\documentclass[12pt]{elsart}
\usepackage{amsfonts}
\usepackage{graphicx}
\usepackage{epsfig}
\usepackage{graphics}

\makeatletter
\newcommand{\row}[1]%
{\mathord{\buildrel{\lower3pt%
\hbox{$\scriptscriptstyle\rightarrow$}}\over #1}}

\newcommand{\dyadic}[1]{\mathord{\dyadic@rrow{#1}}}
\newcommand{\dyadic@rrow}[1]{
\begin{picture}(12,12)(-1,0)
\put(-4,17){\makebox(0,0)[t]{$\scriptscriptstyle\downarrow$}}
\put(-4,17){\makebox(0,0)[l]{$\scriptscriptstyle\longrightarrow$}}
\put(5,0){\makebox(0,0)[b]{$#1$}}
\end{picture}
}

\newcommand{\tr}[2][\,]{\mathrm{tr}_{#1}\left\{#2\right\}}
\newcommand{\bra}[1]{\bigl\langle #1 \bigr|}
\newcommand{\ket}[1]{\bigl| #1 \bigr\rangle}

\newcommand{\PRA}[3]{Phys.\ Rev.\ A  \textbf{#1}, #2 (#3)}
\newcommand{\plA}[3]{Phys.\ Lett.\ \textbf{A#1}, #2 (#3)}
\newcommand{\JMO}[3]{J. Mod.\ Opt.\ \textbf{#1}, #2 (#3)}
\newcommand{\JOB}[3]{J. Opt.\ B  \textbf{#1}, #2 (#3)}
\newcommand{\Nat}[3]{Nature  \textbf{#1}, #2 (#3)}

\begin{document}
\begin{frontmatter}
\title
{More Efficient Purifying scheme via Controlled- Controlled NOT
Gate}
\author{ N. Metwally $^{1}$and A.-S. Obada $^{2}$}
\address{$^1$ Math. Dept., Faculty of Science, South Valley University, Aswan,
Egypt.}
\address{ $^2$ Math. Dept., Faculty of Science, Al-Azhar
University, Cairo, Egypt.}
\thanks{e. mail:Nasser$\_$ Metwally@hotmail.com}

\begin{abstract}
A new modified  version of the Oxford purification protocol is
proposed.
 This  version  is based on the controlled-controlled NOT gate
instead of controlled NOT in the original one. Comparisons between
the results of the new version and the original and an earlier
modification are given. It is found  that the new version
converges faster and consumes fewer initial qubit pairs of low
fidelity per final qubit pair of high fidelity.
\end{abstract}
\end{frontmatter}


\section{Introduction}
 It is well known that to perform most of the quantum information schemes
 efficiently, one requires maximally entangled states to be performed.
However in  reality, it is mandatory to consider the effect of the
decoherence. In this situation, the maximally entangled states
turn into partially entangled or product states. Therefore, these
schemes my not be realized faithfully. Thus purifying these states
is of a great importance in the context of quantum information.
The entanglement purification, that is often required, distills a
small number of strongly entangled pairs of qubits from a large
number of weakly entangled pairs. This can be achieved  by using
local quantum operations, classical communication and
measurements. Bennett et. al have proposed the first entanglement
purification scheme  which is called the IBM protocol \cite{ben}.
This protocol assumes that the initial states are of Werner type
state. The initial density state for the IBM protocol is given by:
$\rho=x\ket{\psi^-}\bra{\psi^-}+\frac{1-x}{4} I$
which means that an $x$ fraction of the singlet state
$\ket{\psi^-}=\frac{1}{\sqrt{2}}(\ket{01}-\ket{10})$
 \cite{Wer} and the rest is randomized .
 Another  standard protocol has been  proposed
by Deutsch et al. and it is called  the Oxford protocol
\cite{Deut}. This protocol is designed for quantum cryptography.
The initial density operator for this protocol takes the form:
$\rho_{Ox}=A\ket{\phi^+}\bra{\phi^+}+B\ket{\psi^-}\bra{\psi^-}+
C\ket{\psi^+}\bra{\psi^+}+D\ket{\phi^-}\bra{\phi^-}$
where
$\ket{\phi^{\pm}}=\frac{1}{\sqrt{2}}(\ket{00}\pm\ket{11})$ and
$\ket{\psi^{\pm}}=\frac{1}{\sqrt{2}}(\ket{01}\pm\ket{10})$ are the
Bell basis. Since then many theoretical \cite{Pan} and
experimental\cite{Wpan} schemes have been proposed.

The standard protocols \cite{ben,Deut} are based on the controlled
NOT operations, CN and Bell state measurements. The IBM protocol
has been improved by Feng et al \cite{Feng}, where they considered
the controlled-controlled NOT operation (CCN) instead of CN. They
showed that the new version is more efficient than the original
one. Because in order to get a final state of certain fidelity
from pairs of the same fidelity, the successful purification
probability is larger and the amount of resources minimally
consumed is less. Also the Oxford protocol has been improved by
Metwally, where in his version the usual local unitary
transformations  are only performed when they are really helpful
\cite{Metwally}. For some initial states  this modification makes
the new version converges faster, more efficient and needs fewer
operations. For some other initial states the two versions are
equivalent.

In this {\it contribution}, we study the Oxford protocol in the
dynamical variables i.e., Pauli's operators, $\sigma_i$ and
$\tau_i$, where $i=0,x,y,z$ for the first and
 the second qubit respectively.
In this current version we consider the CCN operation instead of
the CN. We see that the modification makes the suggested version
more efficient.
 In section $2$,we describe briefly the Oxford protocol Ox$_1$ and its
first modification Ox$_2$\cite{Metwally}. Also we achieve the CN
operation using the dynamical variables. Section $3$ is devoted to
a study of the second modified version, Ox$_3$. A comparison of
the three protocols is discussed in section $4$.
 Finally we summarize our results in section $5$.

\section{The Oxford protocol, Ox$_1$ and the first modified version, Ox$_2$}
Assume that Alice and Bob are given an ensemble of the so called
generalized Werner states or self transposed states \cite{EM}, or
simply Bell-diagonal states \cite{ben}and they are asked to use
the Oxford protocol to purify them. The users pick two pairs of
the form
\begin{equation}\label{eq:GW}
\rho^{(1)}=\frac{1}{4}(1+c_x\sigma_x^{(1)}\tau_x^{(1)}
-c_y\sigma_y^{(1)}\tau_y^{(1)}+c_z\sigma_z^{(1)}\tau_z^{(1)})
\end{equation}
where
\begin{equation}\label{eq:ord}
1\ge|c_x|\ge|c_y|\ge |c_z|\ge 0
\end{equation}
 the order being a matter of convention.
 These coefficients $c_i,i=x,y,z$ are related to the coefficient of
 the original Oxford protocol by the following relations:
\begin{eqnarray}
c_x&=&A-B+C-D
\nonumber\\
c_y&=&A-B-C-D
\nonumber\\
c_z&=&A+B-C-D
\end{eqnarray}
The initial fidelities of states $(\ref{eq:GW})$ are given by:
\begin{equation}
F_1=\tr{\rho^{(1)}\rho^{(1)}_{\phi^+}}=\frac{1}{4}(1+c_x+c_y+c_z)
\end{equation}
where
\begin{equation}\label{eq:Bell-1}
\rho^{(1)}_{\phi^=}=\frac{1}{4}(1+\sigma_x^{(1)}\tau_x^{(1)}
-\sigma_y^{(1)}\tau_y^{(1)}+\sigma_z^{(1)}\tau_z^{(1)}).
\end{equation}

\vspace{0.2 in}
\begin{table}[b!]
 \begin{center}
\begin{tabular}{c|cccc}
\hline\hline
&$I^{(2)}$&$\sigma_x^{(2)}$&$\sigma_y^{(2)}$&$\sigma_z^{(2)}$\\
\hline
$I^{(1)}$&$1$&$\sigma_x^{(1)}\sigma_x^{(2)}$&$\sigma_y^{(1)}\sigma_x^{(2)}$&$\sigma_z^{(1)}$\\
$\sigma_x^{(1)}$&$\sigma_x^{(2)}$&$\sigma_x^{(1)}$&$\sigma_y^{(1)}$&$\sigma_z^{(1)}\sigma_x^{(2)}$\\
$\sigma_y^{(1)}$&$\sigma_z^{(1)}\sigma_y^{(2)}$&$\sigma_y^{(1)}\sigma_z^{(2)}$&-
$\sigma_x^{(1)}\sigma_z^{(2)}$&$\sigma_y^{(2)}$\\
$\sigma_z^{(1)}$&$\sigma_z^{(1)}\sigma_z^{(2)}$&$-\sigma_y^{(1)}\sigma_y^{(2)}$&
$\sigma_x^{(1)}\sigma_y^{(2)}$&$\sigma_z^{(2)}$\\
 \hline
 \end{tabular}
 \vspace{0.1in}
 \caption{Bilateral CN operation between the two qubits which is defined by
   $\sigma_{\mu}^{(1)}$ and $\sigma_{\mu}^{(2)}$. The same table is applied to
   the two qubits $\tau_{\mu}^{(1)}$ and $\tau_{\mu}^{(2)}$.}
\end{center}
\end{table}

Assume that Alice and Bob want to purify their pairs by using the
{\it first} modified Oxford protocol Ox$_2$. To achieve this aim,
they perform the Bilateral CN operation, see table 1, on the
pairs, followed by measuring the target qubit(second pair) in the
computational basis. They measure the z components of the target
spin $\sigma_z^{(2)}$ and $\tau^{(2)}_z$. They keep those first
pairs for which they get the same results for the measurements and
discard the others. After one step purification, they get
\begin{equation}\label{eq:ox2}
\rho_{{Ox}_2}=\frac{1}{4}\left[
1+\frac{c_xc'_x+c_yc'_y}{1+c_zc'_z}\sigma_x\tau_x-
\frac{c_xc'_y+c_yc'_x}{1+c_zc'_z}\sigma_y\tau_y
\frac{c_z+c'_z}{1+c_zc'_z}\sigma_z\tau_z\right]
\end{equation}
this is another new Bell state with fidelity,
\begin{equation}
F_{Ox_2}=\frac{1}{4P_1}\left[(1+c_z)^2+(c_x+c_y)^2\right]
\end{equation}
where $P_1=\frac{1}{2}(1+c_z^2)$, is the probability that Alice
and Bob obtain coinciding outcomes in the measurement of target
pair\cite{Metwally}. Now Alice and Bob have a new ensemble
described by (\ref{eq:ox2}). If this ensemble does not obey the
ordering required by (\ref{eq:ord}), then Alice and Bob use local
rotations to bring the state into the wanted form. These are
rotations by $\pi/2$ about the $x,y$ or $z$ axis, namely
\begin{equation}\label{eq:unit}
U_{12j}=e^{i\pi(\sigma_j-\tau_j)}\quad \mbox{and}\quad j=x,y,z.
\end{equation}
In fact it is only necessary to ensure that $|c_z|$ is smaller
than $|c_x|$ and $|c_y|$; the relative size of $|c_x|$ and $|c_y|$
does not matter. In the standard protocol, Ox$_1$ \cite{ben},
Alice and Bob perform the transformation (\ref{eq:unit}) in $x-$
direction on all pairs. This operation changes the positions of
$c_y$ and $c_z$. After applying the Bilateral CN operation and
measuring the target qubits, Alice and Bob get a new ensemble with
fidelity
\begin{equation}
F_{Ox_1}=\frac{1}{4P_2}\left[(1+c_y)^2+(c_x+c_z)^2\right]
\end{equation}
where $P_2=\frac{1}{2}(1+c_y^2)$ is the probability that Alice and Bob's measurements
are the same.
\section{The second modified protocol, Ox$_3$}
Before performing this protocol, one needs to describe the controlled-controlled Not, CCN. It is a three qubit gate, two of them are used as a control qubit and the third is a target. Mathematically it takes the following form,
\begin{eqnarray}\label{eq:CCN}
CCN&=&\frac{1}{4}\biggl[(1+\sigma_z^{(1)})(1+\sigma_z^{(2)})\sigma_x^{(3)}+
(1+\sigma_z^{(1)})(1-\sigma_z^{(2)})
\nonumber\\
&&+
(1-\sigma_z^{(1)})(1+\sigma_z^{(2)})+
(1-\sigma_z^{(1)})(1-\sigma_z^{(2)})\biggr]
\end{eqnarray}
The target qubit pairs change only when the two control qubits are
in the state $1$. In a generic form one can write its effect as
$CCN\ket{abc}=\ket{ab,c\oplus a.b}$. As an example it transforms
the vector $\ket{\phi_1^+\phi^+_2\phi^-_3}$ to
$\frac{1}{2}[\ket{\phi_1^+\phi^+_2\phi^-_3}+
\ket{\phi_1^-\phi^+_2\phi^-_3}+ \ket{\phi_1^+\phi^-_2\phi^-_3}-
\ket{\phi_1^-\phi^-_2\phi^-_3}]$. It is possible to consider the
effect on the dynamical variables, but its form is too complicated
to be written in this text.

Assume that Alice and Bob are given an ensemble of identical pairs
in the form of equation (\ref{eq:GW}). To perform the second
modified version of the Oxford protocol Ox$_3$, they pick three
pairs and both of them perform the transformation (\ref{eq:unit})
on all pairs. After applying the CCN operations , they measure the
target qubits in the projection
$\ket{\phi^+}_{A_1A_2}\bra{\phi^+}\otimes\ket{\phi^+}_{B_1B_2}\bra{\phi^+}+
\ket{\phi^-}_{A_1A_2}\bra{\phi^-}\otimes\ket{\phi^-}_{B_1B_2}\bra{\phi^-}$,
where $A_1A_2$ and $B_1B_2$ stands for the first and the second
qubits for Alice and Bob respectively. They will keep those pairs
for which the  measurement results coincide. These pairs are used
for the second round. After one step purification they get a new
ensemble of states,
\begin{equation}
\rho_{Ox_3}=\frac{1}{4}(1+c_x^{new}\sigma_x\tau_x+
c_y^{new}\sigma_y\tau_y + c_z^{new}\sigma_z\tau_z)
\end{equation}
with a fidelity
\begin{equation}\label{eq:Fid3}
F_{Ox_3}=\frac{1}{2P_3}\left[(A^2+D^2)A+(B^2+C^2)C\right]
\end{equation}
and,
\begin{eqnarray}
C_x^{new}=A_{new}-B_{new}+C_{new}-D_{new}
\nonumber\\
C_y^{new}=A_{new}-B_{new}-C_{new}-D_{new}
\nonumber\\
C_z^{new}=A_{new}+B_{new}-C_{new}-D_{new}
\end{eqnarray}
where
\begin{eqnarray}
A_{new}&=&F_{Ox_3},\quad B_{new}=\frac{1}{N_3}\left[(A+C)DB\right]
\nonumber\\
C_{new}&=&\frac{1}{2P_3}\left[A^2D+(B^2+C^2)A+D^2C\right]
\nonumber\\
D_{new}&=&\frac{1}{P_3}\left[AD^2+CB^2\right]
\end{eqnarray}
and $P_3$ the probability of success
\begin{equation}
P_3=\frac{1}{2}\biggl[A^3+(3B^2+D^2)C+(3A+D)AD
+2(A+C)BD+C^3\biggr]
\end{equation}
Now assume that the pairs in the initial ensemble are not identical i.e
 with different $c_i, i=1,2,3$  and with different fidelities.
 Alice and Bob want to perform the Ox$_3$.
 To achieve this goal, they pick three pairs of the form
\begin{equation}\label{eq:diff}
\rho=\frac{1}{4}(1+c_{xi}\sigma_x\tau_x
-c_{yi}\sigma_y\tau_y+c_{zi}\sigma_z\tau_z), \quad i=1,2,3
\end{equation}
They perform all the required local operations as described above.
After one iteration, they will get a new ensemble, with a new
fidelity, to  be used for the second round. In this case the
fidelity is given by
\begin{equation}
F=\frac{1}{2P_d}\left[(A_1A_2+D_1D_2)A_3+(B_1B_2+C_1C_2)C_3\right]
\end{equation}
where $P_d$ is the probability of success,
\begin{eqnarray}
P_d&=&\frac{1}{2}\biggl[A_1A_2(A_3+D_3)+(B_1B_2+C_1C_2+D_1D_2)(A_3+C_3)
\nonumber\\
&& +2(A_1D_2+D_1A_2+B_1C_2+C_1B_2)(B_3+D_3)\biggr]
\end{eqnarray}

\begin{figure}
  \begin{center}
\includegraphics[width=35pc,height=18pc]{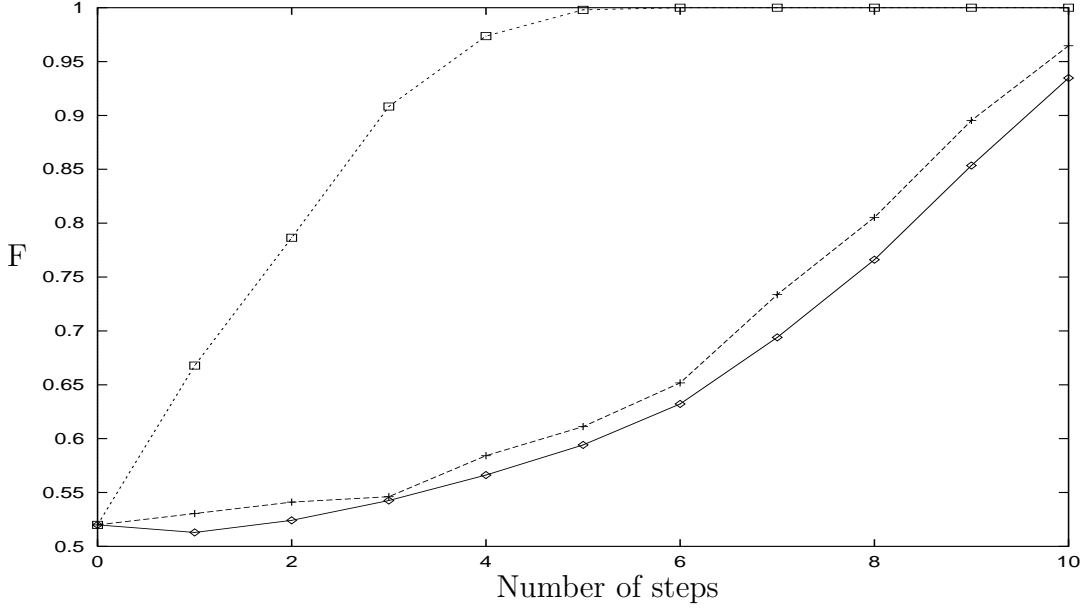}
  \put(-230,-10){Number of steps}
\put(-415,115){F}
  \caption{The Fidelity of the purified pairs, the dot line represents the Ox$_3$,
  the dashed line for Ox$_2$ and the solid line for the original protocol Ox$_1$.
  The initial fidelity of the pairs, $F=0.52$. }
  \end{center}

\end{figure}

\section{Discussion}
To study the efficiency of the  protocols: the original one
Ox$_1$,
 the first modified one, Ox$_2$ and the second modified one Ox$_3$,
  we consider an ensemble of identical states given by (\ref{eq:GW}).
  Lat us assume that this ensemble   has an initially  fidelity, $F=0.52$.
In Fig.$(1)$, we plot the fidelity of the surviving states after
each step as a function of the number of iteration. As a first
remark, the fidelity of the purified state using the Ox$_2$ and
Ox$_3$ after one step  is larger than that obtained using the
standard protocol Ox$_1$.
 It  decreases for the Ox$_1$ and then increases, but for the modified
 ones it  always increases. Concerning the fidelity after one step obtained
 by using Ox$_3$ it is much larger that obtained from Ox$_2$.
Since the fidelity in each next step depends on the previous one,
the final state in the modified protocols converges faster than in
the standard one. Also for the Ox$_3$ it converges much faster
than Ox$_2$. The number of steps needed in the modified protocols
are fewer than those in the standard one. In table $(2)$ there is
a comparison of the three protocols. It is clear that, in order to
get a state with a fidelity  $\simeq 0.8$, one needs $9$ steps for
the Ox$_1$ and $8$ steps for the Ox$_2$ and $3$ steps for the
Ox$_3$. The probability of success in each step, for the original
protocol is larger than the modified ones.
\begin{table}[htp]
\begin{center}
\begin{tabular}{c|c|c|c}
\hline
 protocol&Fidelity&number of iterations&consumed pairs\\
 \hline
Ox$_1$&0.853&9&256\\
Ox$_2$&0.805&8&128\\
Ox$_3$&0.843&3&9 \\ \hline
\end{tabular}
  \caption{Comparison between the Ox$_1$ and its modified versions Ox$_2$ and Ox$_3$. }
\end{center}
\end{table}
To complete comparing the efficiency between the three protocols,
 we have to examine the consumed pairs in each protocol.
 We know that the users in each step needs two pairs for the Ox$_1$
 and Ox$_2$ protocol while three pairs for the Ox$_3$ are needed.
 In each success step, they consume one pair for the Ox$_1$ and Ox$_2$
 but two pairs for the Ox$_3$. Table $(2)$ shows the number of the consumed
 pairs in each protocol to get a certain  fidelity. From this table it is clear that
 the resource consumed in the modified versions are fewer than the original one.

\section{Summary}
The new proposed version Ox$_3$ is more efficient than Ox$_1$ and
Ox$_2$ for the following reasons:(i) The final state converges
faster than the two other protocols.
 (ii) The steps needed to get a state with a certain fidelity are fewer
  than those  needed if the  other protocols are used.
  (iii)The consumed sources are much fewer.
  (iv) The new version, Ox$_3$ is efficient for any set of initial
  states, while the Ox$_2$ and Ox$_1$ are the same for the set of
  Werner states.
 However, we would point out to the only disadvantage of this protocol namely:
  the successful purification probability is
 smaller than the original one, however the Ox$_3$ is much better
 than Ox$_2$.

 {\bf Acknowledgments}

 We thank the referees for their objective comments that improved
 the text on many points.


\bigskip


\begin{thebibliography}{nas}
\bibitem{ben}
C. H. Bennett, G. Brassard, S. Popescu, B. Schumacher, J. A.
Smolin and W. K. Wottor, Phys. Rev. Lett.{\bf 76},722 {(1996)}; C.
Bennett, D. P. DiVincenzo, J. A. Smolin and W. K. Wootter,
\PRA{54}{3824}{1996}.

\bibitem{Wer}
R. F. Werner, \PRA{40}{4277}{1989}; A. Peres, Phys. Rev. Lett.{\bf
77}, {1413} {(1996)}.
\bibitem{Deut}
D. Deutsch, A. Ekert, R. Jozsa, C. Macchiavello, S. popescu and A.
Sanpera, Phys. Rev. Lett {\bf 77},2818{(1996)}; C. Macchiavello,
\plA{246}{385}{1998}.
\bibitem{Pan}
J. -W. Pan et al., Phys. Rev. Lett. {\bf 80}, 3819
{(1998)};Bao-Sen Shi, Yun-Kun- Jiang and Guang-Ca Guo
\PRA{62}{054301}{200}; J. -W. Pan, C. Simon, C. Brakner and A.
Zeiliner, Nature{\bf 410}(2002); Liu Ye, Chun-Mei Yao and Guang-
Can Gu, \JOB{3}{21}{2002} and W. Dur and H.-J. Briegel Phys. Rev.
Lett. {\bf 90}, 067901 {(2003)}.
\bibitem{Wpan} J. W. Pan, S. Gasparoni, R. Ursin, G. Weihs and Zilinger, \Nat{423}{417} {2003}.
\bibitem{Feng}
Xun-Li Feng, Shang-Qing Gand Zhi-Zhan Xu,\plA{271}{44}{2000}.
\bibitem{Metwally}
N. Metwally, \PRA{66}{054302}{2002}.
\bibitem{EM}
B.-G. Englert and N. Metwally \JMO{47}{221}{2000}; B.-G. Englert and N. Metwally, {\it Mathematics of Quantum computation}, edit by R. Brylinski and G. Chen (CCR, Boca Raton, FL, 2002), pp. 25-75.

\end{thebibliography}
\end{document}